\documentstyle[galley]{mn}
\begin{document}

\title[LYMAN ALPHA FOREST TOWARDS B2 1225+317]
{LYMAN ALPHA FOREST TOWARDS B2 1225+317}

\author[P. Khare et al]{Pushpa Khare$^1$, R.Srianand$^2$, D.G.York$^3$, R.Green$^4$,
\and D.Welty$^3$, Ke-Liang Huang$^5$, and J.Bechtold$^6$\\
$^1$ Physics Department, Utkal University, Bhubaneswar,751004, India\\
$^2$ IUCAA, Post bag 4, Ganeshkhind, Pune, 411007, India\\ $^3$
University of Chicago, 5640 S.Ellis Ave., Chicago. IL60637 USA\\ $^4$
National Optical Astronomical Observatories, Tuscon, AZ 85726-6732
USA\\ $^5$ Nanjing Normal University, Department of Physics, 122
Ninghai Road, Nanjing, Jiangsu 210024, China\\ $^6$ Steward
Observatory, University of Arizona, Tuscon AZ 85721 USA\\}
\pagerange{000,000}
\pubyear{1996}

\maketitle

\begin{abstract}
We present observations of the Lyman alpha forest towards B2 1225+317
taken at a resolution of 18 km s$^{-1}$. A clean sample of Lyman alpha
forest lines is extracted after a careful profile fitting analysis and
removal of absorption lines of heavy elements. The sample is analyzed
for statistical properties. Eighty percent of the column densities are
$<$ 10$^{14}$ cm$^{-2}$. A single power law is inconsistent with the
column density distribution and a steepening/break in the distribution
is indicated. The average velocity dispersion parameter is 29.4 km
s$^{-1}$. We find 3$\sigma$  evidence for a correlation between column
density and the velocity dispersion parameter. The correlation,
however, is mainly due to narrow lines and weakens to 1.2$\sigma$ if
lines with velocity dispersion parameter smaller than 20 km s$^{-1}$ are
excluded. An excess of line pairs with velocity separation $\le$ 100 km
s$^{-1}$ over the expected number is found. 
\end{abstract}


\section{INTRODUCTION}

Lyman alpha forest lines in the spectra of QSOs are generally believed
to be produced by intergalactic clouds (Sargent etal 1980), though some
evidence exists for their being associated with galaxies (Lanzetta etal
1995; Le Brun et al, 1996; Bowen et al 1996). An understanding of the
properties of these lines is crucial not only for the understanding of
their nature, but also for the understanding of the formation and
evolution of structure in the Universe as well as that of the physical
conditions existing in the universe at redshifts up to 5. These lines
are in fact the only available probes of the intergalactic medium in
the Universe at these times. A large amount of observational data for
these lines, taken at intermediate resolution (FWHM $\simeq 1.0 \AA$),
have been compiled over the years. Because the Lyman alpha lines are
numerous, line blending in the intermediate resolution data  poses a
severe handicap. It is therefore necessary to have high resolution
observations of these lines to understand their true properties such as
the column density, velocity dispersion distributions and clustering.
Such data are becoming available and to date observations are available
for about 15 quasars at resolutions ranging from 6 km s$^{-1}$ to 35 km
s$^{-1}$. These observations have already provided interesting
information about the Lyman alpha forest lines. The lines are found to
be clustered on scales of up to 300 km s$^{-1}$. The column density
distribution is found to have a break at $\rm N_{\rm H\; I} \sim
10^{14}\; km\; s^{-1}$. Most importantly some lines have been found to
have velocity dispersion parameters smaller than 20 km s$^{-1}$
(Pettini et al 1990, Srianand and Khare 1994, Kulkarni etal 1996 etc).
Since these properties are statistical in nature, more data is always
welcome to get a better understanding of the line properties.

This paper presents new observations of Lyman alpha forest lines
towards B2 1225+317 at a resolution of 18 km s$^{-1}$ with signal to
noise ratio larger than 10. We have analyzed the data by profile
fitting the observed lines. Heavy element lines are carefully searched
and removed. The statistical properties of the clean sample of Lyman
alpha forest lines are studied.  In section 2, we present details of
the observations; in section 3, we describe the profile fitting
procedure.  In section 4, we discuss the details of the heavy element
line systems reported in the literature for this QSO as well as the
results of the profile fitting analysis for heavy element lines
belonging to these systems that fall in our observed range. In section
5, we present the results of our analysis of the statistical properties
of these lines.  This is followed by conclusions in section 6.
 
\section{observations}

The near UV observations of B2 1225+317 were obtained on the nights of
23 and 26 March, 1987, with the Kitt Peak Mayall 4-meter telescope plus
echelle spectrograph.  The instrumental configuration included the 58.5
l/mm 63 deg echelle grating with the blaze peak centered on the
format.  The 226-1 cross-disperser grating was used in second order
through a copper sulfate filter to allow coverage of orders 96 through
65, or 3130 to 4500 \AA.  The poor seeing on both nights of 3 - 4'' FWHM
and the intermittent clouds on the second night imposed a substantial
throughput penalty to maintain the resolution through the 1 arcsecond
slit. A decker masked the slit to a height of 10 arcseconds, necessary
for separating the orders at their closest spacing at the violet end of
the echellogram.  The spectrograph was rotated between the 3000-second
exposures to orient the slit approximately along the parallactic angle
to avoid selective losses in the far UV.  Each object exposure was
preceded and followed by Th-A argon arc lamp images for interpolation
of the wavelength solution.  The total exposure time on the quasar was
392 minutes on 23 March and 342 minutes on 26 March.  

The spectrograph output was coupled through the "UV Fast" camera to the
Intensified CCD Detector.  That detector consisted of an RCA two-stage
magnetically focused "Carnegie" image tube with a 38 mm cathode, lens
coupled to the "TI3" CCD.  The DQE of the photocathode has a flat peak
at about 18\% from 3500 to 4500 \AA; the efficiency of the cross-disperser
grating falls rapidly longward of 4000 \AA, however, giving reduced S/N
in the redder orders.  For calculation of the S/N, a system gain of 20
data numbers per detected photon was assumed, corresponding to about 80
electrons at the CCD for each output phosphor scintillation.  The
agreement with the S/N ratio derived from local continuum intervals
suggests that this value for the gain is a good estimate.  The slit
afforded 17 km s$^{-1}$ resolution FWHM as measured in single Th-A arc
exposures;  the final combined image from the entire night shows 18
km s$^{-1}$ resolution, as measured from a combination of all the arc
exposures, rebinned identically to the quasar spectra.  The resolution
was limited essentially by the sampling of the slit profile with 2
15-micron pixels; the magnification of the transfer lens between the
image tube and CCD was a factor of 1.66.  Because of the substantial
photon gain, the detector works nearly as a photon counter, in that the
CCD read noise is negligible; exceptions are the significant defects in
the TI3 chip and the fact that ion events are fully amplified in the
integrated image.

\section{FITTING PROCEDURE}

The profile fitting algorithm and analysis are similar to those
described by Kulkarni et al. (1996), except that we also use $\chi^2$
per degree of freedom as a measure of goodness of fit as used by
several authors (Rauch et al. 1992, Christiani et al. 1995). $\chi^2$
and  rms values were calculated over parts of the spectrum free of
absorption lines. While considering the goodness of a profile fit for a
line, we used the values of rms and $\chi^2$ calculated in the line
free regions extending up to 10 $\AA$ on either side of the line. The
minimum number of components necessary to give values of $\chi^2$ per
degree of freedom and rms residual consistent with these values were
used in the fit of each line.
Prior to 1990, ten heavy element line redshift systems had been reported
in the literature for this QSO (York etal 1991). Three additional
systems were found by Steidel and Sargent (1992). Using a search list
similar to that used by Hunstead et al (1986), PK and RS generated a
list of heavy element lines belonging to these systems that would be
present in our spectra. Independently, using an expanded list based on
Morton, York and Jenkins (1988), compiled by DW, DGY made a similar
list of lines of heavy elements that contaminate the lines of the
forest. Each system was analyzed carefully by looking for wavelength
consistency as well as the consistency of line strength for doublet or
multiplet lines of the same species. Details of the heavy element line
systems present and the heavy element lines identified in each of them
are given in next section. Detailed profile fitting and deblending of
Ly $\alpha$ lines, as described by Kulkarni et al (1996), was possible
only for system A, which was studied in detail earlier by Bechtold,
Green and York (1987); for system H, for which the Mg II doublet is
present in our data; and for system I and J, for which C IV lines are
available.  For most other systems no lines were found outside the Ly
$\alpha$ forest in our spectra and no multiplets were found inside the
Ly $\alpha$ forest which complied with the consistency tests mentioned
above. In some cases, lines were present only at the expected positions
of O I(1304) and C II(1331) lines.  As they are single lines and no
other lines belonging to these systems were present outside the forest,
and as the Ly $\alpha$ lines for these systems, whenever present, are
saturated, it was not possible to deblend any Ly $\alpha$ forest lines
from these lines. We fitted them as Ly $\alpha$ lines. Note that none
of these lines are included in our sample of lines defined below. The
spectrum along with the fitted profiles is presented in figure 1.
 The line positions are tick marked in the figure. For the Ly $\alpha$
lines thus compiled we used a computer programme described by Srianand
\& Khare  (1994) to search for new heavy element line systems.  We do
not detect any new systems.

The list of all lines studied is given in Table 1, which lists lambda;
log N$_{\rm X}$ for species 'X' noted in the last column; the Doppler
parameter, b, for the profile fits; and the redshift of the line. The
notations after the line wavelength, in column 1, are "b", line has
unusual profile due to either bad pixel e.g.  $\lambda3392\AA$,
$\lambda3338\AA$ etc, or line is broad and/or noisy and it's reality is
not certain e.g. $\lambda3382\AA$ ; this selection criterion is
subjective to some extent, however, two of us (PK and RS) independently
arrived at the same set of lines which fall in this category; "d", line
deblended from heavy element line; and "s", evidently, a single
component (a fact used later in this paper). Part of the spectrum with
wavelength smaller than 3400 $\AA$ has S/N value between 2 and 10 while
the part with wavelength larger than 3400 $\AA$ has S/N value between
10 and 20. The completeness limit for our data was determined by
calculating the minimum necessary column density, N$_{\rm H\;I}$, for a
given value of the velocity dispersion parameter such that the line
will be detected at the 5 $\sigma$ level for rms  error of 10 $\%$ (
which is the upper limit to the rms for $\lambda>3400\AA$). These
values are plotted in figure 2 as a continuous line. It can be seen that
lines with b $<$ 100 km s$^{-1}$ could be detected in our spectra
(with $\lambda>3400\AA$) for log N$_{\rm H\; I}$ cm$^{-2}$ $>$ 13.2. For
statistical analysis we have, therefore, compiled a sample, S1, of
lines having log N$_{\rm H\; I}$ cm$^{-2}$ $>$ 13.2 and
$\lambda>3400\AA$, excluding lines within 8 Mpc from the QSO and lines
with unusual profiles (marked with 'b', hereafter LWUPs). The lines
included in S1 are marked with a * in Table 1. Lines marked with '?'
are the LWUPs, which otherwise satisfy the criterion for inclusion in
the Lyman alpha sample. These lines along with S1 form our extended
sample S2.

\section{INFORMATION ABOUT HEAVY ELEMENT LINE SYSTEMS}

\noindent{\bf System A : (z= 1.79)}

This system has been studied in high resolution ($\sim\;10\;{\rm
km\;s^{-1}}$) by York et al. (1984) and by Bechtold, Green and York
(1987).  They observed C IV, Fe II, Mg II and Si II lines which were
fitted with up to 12 components with redshifts in the range
1.7933-1.7975. Our spectrum covers 1136 - 1435 $\AA$ in the rest frame
of the system. The Si II 1190, 1193 and Ly $\alpha$ lines are distorted
due to pixel defects or cosmic ray hits. These deleterious effects
prohibit deblending of Lyman alpha lines from the Si II lines. The main
Lyman alpha line is too saturated to allow fitting of the components
seen in the heavy element lines at these redshifts. Detailed
photoionization modeling for this system has been considered by
Bechtold, Green and York(1987).  We are able to fit most lines with 7
components with redshifts close to the redshifts obtained by Bechtold,
Green and York (1987).

\noindent{\bf System B: (z = 1.6251)}

Wilkerson et al. (1978) observed C IV , Si IV, C II, Si II and Ly
$\alpha$ lines belonging to this system. Mg II lines were not detected
by Steidel and Sargent (1992). Young, Sargent \& Boksenberg (1982)
detected C IV lines. Our spectrum covers 1211 - 1530 $\AA$ in the rest
frame of this system. A line is present at the expected position of Si
IV 1402. However no line is observed at the expected position of 1393.
Based on the upper limits from the absence of this line, we conclude
that the line at the position of Si IV 1402 is a Ly $\alpha$ forest
line and any Si IV line if present can not alter the deduced parameters
of this Ly $\alpha$ line significantly. Si II (1260) and  C II (1334)
lines are possibly present. In our low quality data near 3180 $\AA$
(not shown here), we find Si III $\lambda$1206 (W$_\lambda$ = 0.92$\pm
0.15 \AA$,1 $\sigma$) and Lyman alpha (W$_\lambda$ = 3.94$\pm 0.23
\AA$). No other heavy element lines are present. As Mg II and Fe II
lines have not been observed for this system, the reality of C II and
Si II lines is doubtful. Also the Ly $\alpha$ line lies in very low S/N
region. We have therefore not attempted to deblend any Ly $\alpha$
lines from these lines, but have excluded these lines from Ly $\alpha$
line sample.

\noindent{\bf System C: (z = 1.8865)}

Young, Sargent and Boksenberg (1982) detected only C IV lines belonging
to this system. The displayed spectrum covers 1101-1391 $\AA$ in the
rest frame of this system. Si IV is not present in our spectrum at
slightly longer wavelength than the region discussed here. We find the
Ly $\alpha$ line belonging to this system.  This can be fitted with 5
components, the redshifts of which differ slightly from the values
quoted by Steidel \& Sargent (1992). Lines of O I and C II are possibly
present.  These lines are removed from our Ly $\alpha$ line sample. No
other heavy element line is found for this system.  We note that the
profile of the shortward wing of the Lyman alpha is fit by a damped
profile, with N$_{\rm H I } = 2.5\times 10^{17}$ cm$^{-2}$. The clean
wing yields an upper limit for deuterium, N$_{\rm D I }\;\le\;1.6\times
10^{13}$, giving, 
N$_{\rm D I }$/N$_{\rm H I }\;\le$ 6.4 10$^{-5}$.

\noindent{\bf System D: (z = 1.8963)}

Wilkerson et al. (1978) detected Ly $\alpha$ and C IV lines for this
system.  Young, Sargent and Boksenberg (1982) reported the detection of
C IV lines alone. Our spectrum covers 1098 -1386 $\AA$ in the rest
frame of this system. We find Ly $\alpha$ line belonging to this
system: this line is blended with Si II (1260) of system A. The H I
portion is fitted with 3 components, the redshifts of which differ
slightly from the values quoted by Steidel \& Sargent (1992).
Components of  C II are possibly present. These lines are excluded from
our Ly $\alpha$ line sample. No other heavy element line is found for
this system.

\noindent{\bf System E: (z = 2.1103)}

This system was reported by Wilkerson et al (1978). However the system
is doubtful as Ly $\alpha$ is very weak. Young, Sargent and Boksenberg
(1982) and Steidel and Sargent (1992) did not identify this system. Our
spectrum covers 1022 - 1291 $\AA$ in the rest frame of this system. We
also do not find any line belonging to this system and thus conclude
that this system is not real.

\noindent{\bf System F: (z = 1.3582)}

Wilkerson et al. (1978) found only marginal evidence for the presence
of this system. Steidel \& Sargent (1992) did not find Mg II for this
system. They also reported the absence of this system in their
unpublished high resolution data. Our spectrum covers 1348 - 1702 $\AA$
in the rest frame of this system. We do not find C IV lines with proper
redshift matching and line strengths. No other heavy element lines are
detected.  Thus we conclude that this system is not real. We note,
however, that the first three lines in Table 1 could be Si IV
$\lambda$1393 . Matching components for Si IV $\lambda$1402 are
present. It is unusual to find Si IV without C IV (but see Khare, York
and Green 1989).

\noindent{\bf System G: (z = 2.1197)}

Wilkerson et al (1978) found Ly $\alpha$, Ly $\beta$, N I, C IV and O
VI lines belonging to this system. Young, Sargent and Boksenberg (1982)
found weak C IV doublet. Our spectrum covers 1019 - 1287 $\AA$ in the
rest frame of this system.  We identify Ly $\alpha$, Ly $\beta$ and
possibly N V lines. Other heavy element lines are not present. In
particular the N I lines suggested earlier are definitely not present,
being confused with nearby Lyman alpha lines and a possible line of C
II from another system.

\noindent{\bf System H: (z = 0.363)}

This system was reported in Steidel and Sargent (1992) based on their
unpublished high resolution data. Our spectrum covers 2332 -2934 $\AA$
in the rest frame of this system. We identify the Mg II doublet, and
deblend the associated Lyman alpha lines.  Mg I is probably present and
Fe II $\lambda$2600 is possible, but the feature does not match the Mg
II profile well. 

\noindent{\bf System I: (z = 1.2252) }

This system with two components was reported in Steidel and Sargent
(1992) based on their unpublished high resolution data. However, they
did not find Mg II, Mg I and Fe II lines belonging to this system. Our
spectrum covers 1429 - 1804 $\AA$ in the rest frame of this system.  We
identify only a weak C IV doublet belonging to this system.  Lyman
alpha lines are separable at our resolution, but are situated so as to
have simulated a stronger, more complex profile, in data of slightly
lower resolution. 

\noindent {\bf System J: (z = 1.4290)}

This system was reported in Steidel and Sargent (1992) based on their
unpublished high resolution data. They did not find Mg II, Mg I and Fe
II lines belonging to this system. Our spectrum covers 1309 - 1653
$\AA$ in the rest frame of this system.  We identify only a C IV
doublet belonging to this system. Si IV is absent.

\section{RESULTS OF STATISTICAL ANALYSIS}

\subsection{N$_{\rm H\; I}$ distribution}

Our sample contains single lines and lines fitted with multiple
components as well as lines deblended from heavy element lines. It is
possible that the deblending  procedure may not be ideal and might
artificially introduce some bias in the derived line parameters in the
data. In order to check this possibility we performed several KS
tests. The probability that the column densities for the set of lines
which were fitted with single components and the set of lines which
were fitted with multiple components belong to the same parent
population is $\sim$ 0.18. This indicates that the fitting procedure
does not introduce any bias in the N$_{\rm H\; I}$ distribution. The
probability that N$_{\rm H\; I}$ values for lines lying in the region
of the spectrum with S/N $>$ 10 and S/N $<$ 10 are drawn from the same
parent population is 0.86 indicating that the S/N does not affect the
column density estimates significantly. 

All the  Ly $\alpha$ lines in our list have column density between
$5\times 10^{12}$ and $1.2\times 10^{15}$ cm$^{-2}$. We performed maximum
likelihood analysis, for sample S1, to determine the index, $\beta$, of
the assumed power law column density distribution. The resulting values
of $\beta$ for various lower limiting column densities, N$_{\rm H\;
I}^{\rm min}$ along with the Kalmogoaroff Smirnoff (KS) probability, ${\rm
P_{KS}}$, that the observed distributions are well represented by the
power law are given in Table 2. The probability is only 0.088 for log
N$_{\rm H \;I}^{\rm min}$ $\ge$ 13.2, indicating that a single power law
does not provide an acceptable description of the observed column
density distribution. A single powerlaw does provide a good fit for lines
with log N$_{\rm H \;I}^{\rm min}$ $\ge$ 13.4. The values of $\beta$ and
${\rm P_{KS}}$, however, increase with increase in the minimum column
density cutoff, which is consistent with the steepening of the power
law at high column densities as suggested by Petitjean et al (1993) and
Kulkarni et al (1996). The values of $\beta$ for S2 are also
very similar to those in Table 2.

Large samples are always better for determining the statistical
properties. We have combined the lines towards Q1331+170 (sample S2 of
Kulkarni et al.1996, which were observed with the same resolution, have
similar S/N and cover similar redshift range as our sample), towards
1101-26 (Carswell et al 1991) and towards 2206-199 (Rauch et al 1993)
with our sample S1. The last two QSOs have been observed with somewhat
different resolution but cover similar redshift range. The results of
maximum likelihood analysis for this extended sample are also given in
Table 2.  The extended sample has the KS probability of 0.039 for log
N$_{\rm H\; I}^{\rm min}$ =13.2, indicating that a single power law
does not give an acceptable description of the data. A change in the
slope of column density distribution with column density is also
indicated.  Rauch etal (1992) showed in the case of Q0014+813 that a
single power law describes the data well with P$_{\rm KS}$ = 0.51.
Giallongo et al (1993), considering only lines in the spectra of
Q2126-158, got P$_{\rm KS}$= 0.43 for a single powerlaw fit. However,
when lines from these QSOs were combined with lines from 1101-264,
Giallongo et al (1993) showed that a single powerlaw is acceptable with
a probability of only 0.02. They reported this as a signature for the
break in the power law distribution.  Cristiani et al (1995) showed for
Q0055-269 that the probability for a single power law distribution is
0.0007. They attributed the difference in significance from Giallongo
et al's results for Q2126-158 to the small number of lines in that
sample compared with that for Q0055-269. Recently Giallongo et al
(1996) have confirmed the break with an extended high resolution
sample. They suggest that the line blanketing effect due to the high
column density lines that conceals weak lines, though appreciable at
high redshifts, may not be the cause for the observed break. We have
performed a double power law fit to our extended sample. The column
density distribution along with the power law fits are shown in figure
3. The slopes of the power laws for N$_{\rm H\; I}$ $\le$ 14.0 and
N$_{\rm H\; I}$ $\ge$ 14.0 being -1.15 and -2.05 respectively.  Giallongo
et al (1996), for 1100 Lyman alpha lines in the redshift range 2.8-4.1
have obtained the values for the corresponding slopes to be -1.4 and
-1.8, which are somewhat different than our values.

Thus either, as suggested by Cristiani et al (1995), there exists a
column density cutoff at about 10$^{14}$ cm$^{-2}$ and the higher
column density lines are blends of weaker lines which could not be
deblended due to the finiteness of the resolution and signal to noise
or the break in the column density distribution is due to some physical
process. Hu etal (1995) also find a steepening of the powerlaw beyond
the column density of $3\times10^{14}$ cm$^{-2}$, in their high
resolution and very high S/N KECK data. It, thus appears that the
steepening of the column density distribution may be due to some
physical effect. In the frame work of the structure formation models,
the break may be a consequence of the onset of Jean's instability
(Meiksin and Madau 1993). While the higher column densities are due to
virialized objects, the lower column density clouds may just reflect
the density fluctuations in the intergalactic medium (Miralda-Escude \&
Rees, 1993). The steeper slope obtained for stronger lines for our
sample compared to the slope obtained by Giallongo et al (1996) for
similar lines at higher redshift is consistent with this picture. The
break could possibly be the result of the presence of two populations of
the Lyman alpha clouds as suggested by the KECK observations. While the
high column density lines may be associated with galaxies the lower
column densities may come from truly intergalactic clouds.

If the break in the column density distribution is real and if the
number density evolution index $\gamma$ increases with an increase in
column density (Srianand \& Khare 1994), then one would expect to see
the break more prominently in low redshift samples. More high
resolution, profile-fitted data are required to confirm the existence
and the redshift dependence of the break.  If the break is real, it
will occur at lower column densities in the vicinity of QSOs due
proximity effect and one should see a change in the column density
distribution in the vicinity of the QSOs.  Existing high resolution
data however do not show any such change (Srianand $\&$ Khare, 1996).

\subsection{b-distribution}

The Lyman $\alpha$ lines have b values ranging from 5 to 80 km
s$^{-1}$. A histogram of b-values for S1 is plotted in figure 4. We
have also plotted the histogram for LWUPs and for all the lines with
$\lambda \ge 3400\AA$ without any column density cutoff but excluding
the LWUPs. It can be seen that the lines with column density smaller
than 10$^{13.2} $cm$^{-2}$ have b values smaller than the stronger
lines, which indicates a correlation between the column density and b;
it may also be due to the incompleteness of the sample for column
densities smaller than 10$^{13.2} $cm$^{-2}$. This will be discussed
further in the next section. Also the LWUPs have b values which are
larger than the rest of the lines. The KS probability that these lines
are drawn from the same parent population as the rest of the lines is
only 2$\times 10^{-7}$. The mean and median values for sample S1 are
29.4$\pm$7.9 and 27.66 km s$^{-1}$ respectively. The mean and median
values for sample S2 are 32.1$\pm$9.7 and 30.6 km s$^{-1}$
respectively. The mean and median values for lines in S1, which are
fitted with single components are 27.4$\pm$6.9 and 27.3 km s$^{-1}$
respectively, very close to the values for the whole sample. KS test
shows that the probability that the b values of lines fitted with
single components and lines fitted with multiple components are taken
from the same parent population is 0.65. Thus the fitting procedure
does not introduce any artificial bias in the b-value distribution. The
average b-value is similar for regions of spectra with S/N $>$ 10 and
S/N $<$ 10 and for the lines deblended from heavy element lines
(28.6$\pm$6.1). The KS test shows that the probability that the
b-values for regions of spectra with S/N $<$ 10 and S/N $>$ 10  are
taken from the same parent population is 0.53. The probability that the
b-distribution of lines in B2 1225+317 and in Q1331+170 are drawn from
the same parent population is 0.97.

There is a considerable debate in the literature over the temperature
of Ly $\alpha$ clouds. Pettini et al. (1990) found most b-values for
lines in the spectra of Q2206-199N to be smaller than 22 km s$^{-1}$
and suggested that Ly $\alpha$ clouds are cool having a temperature
between 5000 and 10,000 K. This conclusion was however contradicted by
Carswell et al. (1991) who found an average b-value of 30 km s$^{-1}$
for Q1101-264. Average b-values found for various QSOs studied at high
resolution vary between 27-34 km s$^{-1}$ (Fan and Tytler 1994,
Cristiani et al 1995, Kulkarni et al. 1996). Rauch et al (1993), using
simulated spectra, showed that some of the narrow (b $\le$ 15 km s$^{-1}$)
lines could arise artificially in the spectra due to noise. They
suggested that probably all the narrow lines could be accounted for by
such artificial lines and some unidentified heavy element lines.
Recentely Lu et al (1996) have reached the same conclusion. They found
a lower limit of 15 km s$^{-1}$ to the b values in their data for
Q0000-26. Hu etal (1995) have been able to identify most of the lines
with b $<$ 20 km s$^{-1}$ with metal lines and therefore propose a
cutoff value of 20 km s$^{-1}$. Giallongo etal (1996), however, find
that about 15$\%$ of the lines in their sample have b values between 10
- 20 km s$^{-1}$.

In our sample $\sim 19\%$ of the lines have b-values smaller than 20 km
s$^{-1}$. We have not been able to identify these lines as heavy
element lines. 24$\%$ of the lines fitted with  single components have
b-value $<$ 20 km s$^{-1}$ while 16$\%$ of the lines fitted with
multiple components or deblended from heavy element lines have b$<20$
km s$^{-1}$. As noted above, low column density lines show low b-values
more often compared to high column density lines. While 44$\%$ of the
lines in our sample with log N$_{\rm{H\;I}}$ $<$ 13.5 have b-value
$\le$ 20 km s$^{-1}$, only 8$\%$ of the lines with log N$_{\rm H\; I}$
$>$ 13.5 have b-value  $\le$ 20 km s$^{-1}$.  This is consistent with
the results for other QSOs (Fan \& Tytler, 1994). Another interesting
point to note is that if we consider only clean lines which are fitted
with single components there is no line with b $<$ 20 km s$^{-1}$ for
log N$_{\rm H\; I}$ $>$ 13.5. A similar trend is noted by Giallongo et
al. (1993), who concluded that either such lines are really absent or
they are systematically hidden in blends. corresponding to a cloud
temperature of 24000 K.

An upper limit of 55 km s$^{-1}$ to the b-value for thermally stable
clouds has been obtained by Donahue and Shull (1991). Press and Rybicki
(1993) have pointed out that the high b-values can not be thermal as
otherwise the baryon density contributed by the Ly $\alpha$ clouds will
exceed the upper limit given by the big bang nucleosynthesis. Note that
the b-values for sample S1 are mostly smaller than 50 km s$^{-1}$.

Above results indicate the presence of some correlation between b-value
and the column density, which we discuss in detail in the following
section, or that the stronger lines are unresolved blends of lower
column density components with smaller b-values.

\subsection{N$_{\rm H\; I}$-b correlation}

There is a considerable controversy over the presence of a correlation
between N$_{\rm H\; I}$ and b. Pettini et al (1990) showed a clear
correlation in the case of Ly $\alpha$ lines observed in the spectra of
Q2206-199. They considered only unsaturated unblended lines for the
profile fitting and did not put any completeness limit on N$_{\rm H\;
I}$. This was questioned by Carswell etal (1991) who using a similar
observational setup found no correlation in the case of Q1101-264.
Rauch et al (1993) reanalyzed the Q2206-199 spectra and showed that the
apparent correlation of Doppler parameter and  column density can be
accounted for entirely by biases in the line finding and fitting
procedure due to the finite signal to noise. Invariably all the spectra
show this correlation when the completeness limit to the column density
is not applied. Complete column density limited samples usually do not
show a clear correlation (Rauch et al. 1992, Kulkarni et al 1996.
etc). Recentely Savaglio et al (1996) have shown that at least some of
the N$_{\rm H\; I}$ vs. b correlation may result from the fact that
strong Lyman alpha lines are blends of weaker lines which get
identified only when a simultaneous fit to the Lyman alpha and Lyman
beta lines is performed. N$_{\rm H\; I}$ vs. b plot for our sample is
given in figure 2. The figure also shows the 5$\sigma$ completeness
limit for our sample, which was calculated as described in section 3.

The results presented in the last section indicate the presence of
correlation in our data. Spearman rank correlation test applied to the
whole data without any cutoffs but excluding lines within 8 Mpc of the
QSO shows a 3.62 $\sigma$ correlation with the chance probability of
only 2.2$\times 10^{-4}$. Excluding the LWUPs 
increases the correlation to 5.5 $\sigma$ level. Sample S1 shows a 3.12
$\sigma$ correlation with chance probability of 1.1$\times 10^{-3}$,
while S2 shows a 2.03 $\sigma$ correlation with chance probability of
0.04. The correlation for sample S1 reduces to 1.25 $\sigma$ level if
we exclude lines with velocity dispersion parameter below 20 km
s$^{-1}$. When we consider only clean lines which are fitted with
single components (marked 's' in Table 1, similar to Pettini et al
criteria), with log N$_{\rm H\; I} >$ 13.2, there is a 2.83 $\sigma$
correlation with chance probability 2$\times 10^{-3}$.

We thus confirm the findings of Hu etal (1995) that the correlation
between the column density and velocity dispersion parameter crucially
depends on the presence of Ly $\alpha$ lines with b values smaller than
20 km s$^{-1}$. We have not been able to identify these lines with any
heavy element lines. If these lines are indeed heavy element lines or
have arisen artificially due to the noise then the correlation can be
ruled out. If Ly $\alpha$ lines with small b values do belong to the
forest then the correlation is present but the strength does depend on
the criterion used for selecting the line sample. Our exclusion of
LWUPs from the sample increases the strength of the correlation from
2.03 to 3.12 $\sigma$ level. The correlation found by Pettini etal
(1990) can also be understood as being due to this effect.

\subsection{Clustering properties of Ly $\alpha$ clouds}

The pair velocity correlation between Ly $\alpha$ clouds along the line
of sight has been one of the tools commonly used to study the
clustering properties of Ly $\alpha$ clouds (Sargent et al. 1980).
Webb (1987) showed, in the case of Ly ${\alpha}$ absorption lines
obtained with high resolution spectroscopy, that there is  an excess in
the pair velocity correlation on scales ${\rm \sim 300\;km\;s^{-1}}$.
There are claims and counter claims for the detection of excess on
small velocity scales using high resolution data samples. (Srianand
{\&} Khare 1994, Kulkarni et al 1996, Rauch et al 1992, Pettini et al
1990 etc).
 
We studied the pair velocity correlation in our extended data.
 Sargent et al (1980) used a ramp-shaped function in order to account
for the limited redshift coverage of each spectra.  Instead of taking
any correction function, we calculated the expected number of lines in
various velocity bins taking into account the observable range in the
particular spectra for each Ly $\alpha$ line. The observed number of
pairs in each velocity bin with the expected number and their 2
$\sigma$ (root N) errors are plotted in figure 5 for two different
values of column density cutoffs.  There is a clear excess on the
velocity scales $50-100\;{\rm km\;s^{-1}}$, the distribution being
consistent with the expected distribution for larger velocity
intervals. Note that the deficit in the first bin is the artifact of
blending due to finiteness of resolution. As the correlation function
decreases steeply at high velocity intervals one requires more data to
reduce the random noise in the correlation function, in order to detect
the small amplitude on these velocity intervals. The values of $\xi$
are 1.35 $\pm$0.42 and 3.14 $\pm$1.19 for lines with log N$_{\rm{H\;I}}
\;\ge$ 13.2 and $\ge$13.7 respectively. Dependence of the clustering
amplitude on the strength of the lines in our data is consistent with
the results of Cristiani et al (1995) and Srianand (1996). From the
study of 1100 lines in the redshift interval 2.8-4.1 Cristiani (1996)
finds $\xi$ of 0.2 and 0.6 for weak and strong lines respectively. The
amplitude of the two point correlation function thus increases with
decreasing redshift which is expected in models with hierarchial
clustering. We have studied the correlation of lines, excluding the
lines deblended from the heavy element lines (marked 'd' in Table 1).
The values of $\xi$ are same as the values quoted above showing that of
clustering found here is not due to the deblended lines from the heavy
element lines.

Ostriker, Bajtlik and Duncan (1988) showed that the distribution
function, P(x), of line intervals scaled to the local mean, x, is a
better tool for studying the clustering properties than the pair
velocity correlation. They found a significant excess on the lower
velocity scales in their low resolution sample. The observed line
interval distribution and the expected distribution for the Poisson
distribution of number of clouds along the line of sight, for the
extended sample  are given in figure 6. It is clear from the figure
that there is an excess for small interline spacings over the  expected
value. The excess seems to be more for strong lines, confirming our
pair velocity correlation results.  However KS-test shows the
probability of the maximum difference between the observed and the
predicted distribution to occur by chance is 0.33 and 0.18 for low and
high column density cutoffs respectively. Thus, in spite of a small
excess in the first bin, the distribution is consistent with the
poissonian expectations.

\section{CONCLUSIONS}

We have extracted a clean sample of Lyman alpha forest lines, free from
contamination by heavy element line systems, from the spectra of B2 1225+317,
taken at a resolution of 18 km s$^{-1}$. Lyman alpha forest lines
blended with heavy element lines falling inside the forest have been deblended
whenever possible and included in the sample. The sample consists of
lines with redshifts between 1.7 and 2.2. The results of the analysis
of the statistical properties of this sample can be summarized as
follows.\\
\noindent {1.} The average velocity dispersion parameter of the sample
is 29.4 $\pm$ 7.9 km s$^{-1}$. 19 \% of the lines have b values below
20 km s$^{-1}$. Low b values are more common among weak lines, 44 \% of
lines with log N$_{\rm H\; I}\; < $ 13.5 and 8\% of lines with log
N$_{\rm H\; I}\; > $ 13.5 have b $<$ 20 km s$^{-1}$.

\noindent {2.} A single powerlaw does not give an acceptable fit to the
column density distribution for log N$_{\rm H\; I} \ge 13.2$. For log
N$_{\rm H\; I} \ge 13.4$, a single power law is acceptable, the slope
of the distribution, however, increases with increasing minimum
 column density cutoff, indicating a steepening or break in the power
law. A double power law is fitted to the extended sample of lines,
obtained by combining lines observed towards Q1331+170, Q1101-26 and
Q2206-199. The slopes for log N$_{\rm H\; I} \le 14.0$ and  $\ge 14.0$
are -1.15 and -2.05 respectively.

\noindent {3.} We find evidence for N$_{\rm H\; I}$ - b correlation, it
being significant up to 3.2 $\sigma$ level. The correlation is, however,
mainly due to the narrow (b $\le$ 20 km s$^{-1}$) lines. Exclusion of
these lines weakens the correlation to 1.25 $\sigma$.

\noindent {4.} We find excess of line pairs with velocity splitting
smaller than 100 km s$^{-1}$, the correlation coefficient being 1.35
$\pm$ 0.42 and 3.14 $\pm$1.19 for lines with log N$_{\rm{H\;I}}\; \ge$
13.2 and $\ge$ 13.7 respectively.

\noindent {5.} The interline spacing distribution function shows an
excess over small interline spacings; the excess, however, is not found
statistically significant, the probability of it's occuring by chance
being 0.33.

\section*{acknowledgement}
This work was partially supported by a grant (No. SP/S2/013/93) by the
Department of Science and Technology, Government of India.


%
\noindent{\bf Figure Captions\\}

\noindent {\bf Figure 1.} The observed spectrum of B2 1225+317 for 3278-3942
	   $\AA$ normalized to unit continuum.  The profile fits are
	   shown by continuous lines. Profiles of heavy element lines
	   blended with forest lines are shown as dash-dot lines. Line
	   components are indicated by tick marks.\\

\noindent {\bf Figure 2.}  The plot of log N$_{\rm H\;I}$ vs b.
	  The continuous line indicates the completeness
	   limit of our sample. Crosses represent lines with $\lambda\;
	   \ge 3400 \AA$. Starred triangles represent lines with
	   $\lambda\; \le \; 3400 \AA$.\\

\noindent {\bf Figure 3.}  Column density distribution for the extended
	   sample.  The best fit double power law is also shown. The
	   slopes being -1.15 and -2.05 for log N$_{\rm H\;I}\;\le$ 14.0
	   and $\ge$ 14.0 respectively.\\

\noindent {\bf Figure 4.}  Histogramme of velocity dispersion
	   parameters (b-values) of the Lyman alpha forest lines in
	   Q1225+317. Solid line is for sample S1, dashed line is for
	   LWUPs and dotted line is for all the lines with $\lambda \ge
	   3400 \AA$, excluding LWUPs.\\

\noindent {\bf Figure 5.}  Velocity-pair histogramme for the extended sample.
	   The expected number of pairs as well as the $\pm 2\sigma$
	   values are shown by dashed lines. The upper panel is for
	   lines with log N$_{\rm H\;I}\;\ge\;$ 13.2 while the lower
	   panel is for lines with log N$_{\rm H\;I}\;\ge$ 13.7.\\

\noindent {\bf Figure 6.}  The observed line interval distribution in
	   the extended sample. The expected distribution for
	   Poissonian distribution of clouds is shown by continuous
	   line. The upper panel is for lines with log N$_{\rm
	   H\;I}\;\ge\;$ 13.2, while the lower panel is for lines with
	   log N$_{\rm H\;I}\;\ge$ 13.7.\\

\newpage
\onecolumn
\begin{table}
\caption {Profile fitting results}
\bigskip
\begin{tabular}{lcrcrll} \hline
\multicolumn {1}{c}{$\lambda$}&\multicolumn{1}{c}{log N(X)}&
\multicolumn {1}{c}{b (km s$^{-1}$)}&\multicolumn{1}{c}{z}&
\multicolumn {1}{c}{ID, X}&\multicolumn {1}{c}{}&\multicolumn {1}{c}{Remarks}\\
\hline
  3285.432 &    13.529 $\pm$ 0.06 &     22.09$\pm$ 3.85 & 1.70257 &   H I 1215 & & Po F-Si IV $\lambda$1396  \\
  3286.182 &    13.410 $\pm$ 0.07 &     15.88$\pm$ 4.09 & 1.70319 &   H I 1215 & & Po F-Si IV $\lambda$1396   \\
  3287.105 &    13.626 $\pm$ 0.05 &     32.39$\pm$ 5.07 & 1.70394 &   H I 1215 & & Po F-Si IV $\lambda$1396   \\
  3299.263 &    14.565 $\pm$ 0.08 &     17.91$\pm$ 0.67 & 1.71395 &   H I 1215  &  &  \\
  3299.985 &    14.117 $\pm$ 0.04 &     26.89$\pm$ 1.58 & 1.71454 &   H I 1215  &  &  \\
  3300.800 &    13.526 $\pm$ 0.03 &     22.30$\pm$ 2.56 & 1.71521 &   H I 1215  &  &  \\
  3302.017 &    13.294 $\pm$ 0.03 &     28.62$\pm$ 2.92 & 1.71621 &   H I 1215  &  &  \\
  3305.787b&    13.223 $\pm$ 0.08 &     42.71$\pm$12.01 & 1.71931 &   H I 1215  &  & \\
  3306.614 &    13.290 $\pm$ 0.06 &     21.17$\pm$ 4.18 & 1.71999 &   H I 1215  & & Po F-Si IV$\lambda$1402\\
  3307.411 &    13.456 $\pm$ 0.14 &      9.14$\pm$ 3.73 & 1.72065 &   H I 1215  & & Po F-Si IV$\lambda$1402\\
  3308.402 &    13.449 $\pm$ 0.03 &     37.49$\pm$ 3.96 & 1.72146 &   H I 1215  & & Po F-Si IV $\lambda$1402,\\
           &                      &                     &         &        & & Co B-Si II $\lambda$1260\\  
  3316.489 &    14.305 $\pm$ 0.06 &     32.53$\pm$ 2.98 & 1.72812 &   H I 1215 & & \\
  3317.363 &    13.735 $\pm$ 0.19 &     30.14$\pm$17.18 & 1.72884 &   H I 1215 &  &   \\
  3317.838 &    13.609 $\pm$ 0.19 &     18.09$\pm$ 5.46 & 1.72923 &   H I 1215  & & \\
  3318.684 &    13.830 $\pm$ 0.02 &     31.79$\pm$ 1.99 & 1.72992 &   H I 1215   & &   \\
  3325.052d&    13.654 $\pm$ 0.04 &     36.95$\pm$ 13.38 & 1.73516 &   H I 1215   & &   \\
  3325.685 &    13.776 $\pm$ 0.05 &     31.92$\pm$ 4.53 & 1.79372 &   Si II 1190 &  & \\
  3326.162 &    13.228 $\pm$ 0.18 &     11.33$\pm$ 9.8 & 1.79412 &   Si II 1190  & & \\
  3326.901 &    13.929 $\pm$ 0.09 &     13.74$\pm$ 2.62 & 1.79474 &   Si II 1190  & & \\
  3327.151 &    13.262 $\pm$ 0.22 &     10.71$\pm$ 8.89 & 1.79495 &   Si II 1190  & & \\
  3327.488 &    13.068 $\pm$ 0.27 &      8.93$\pm$ 7.77 & 1.79523 &   Si II 1190  & & \\
  3327.917 &    12.964 $\pm$ 0.71 &     21.89$\pm$ 13.78 & 1.79559 &   Si II 1190  & & \\
  3328.531 &    12.442 $\pm$ 0.12 &     15.56$\pm$ 27.95 & 1.79611 &   Si II 1190  & & \\
  3333.714 &    13.776 $\pm$ 0.05 &     31.92$\pm$ 4.53 & 1.79372 &   Si II 1193  & & \\
  3334.192 &    13.228 $\pm$ 0.18 &     11.33$\pm$ 9.8  & 1.79412 &   Si II 1193  & &\\
  3334.933 &    13.929 $\pm$ 0.09 &     13.74$\pm$ 2.62 & 1.79474 &   Si II 1193 &  &\\
  3335.184 &    13.262 $\pm$ 0.22 &     10.71$\pm$ 8.89 & 1.79495 &   Si II 1193 &  &\\
  3335.521 &    13.068 $\pm$ 0.27 &      8.93$\pm$ 7.77 & 1.79523 &   Si II 1193 &  &\\
  3335.951 &    12.964 $\pm$ 0.71 &     21.89$\pm$ 13.78 & 1.79559 &   Si II 1193 &  &\\
  3336.566 &    12.442 $\pm$ 0.12 &     15.56$\pm$ 27.95 & 1.79611 &   Si II 1193 &  &\\
  3338.096b&    13.599 $\pm$ 0.13 &    103.54$\pm$45.77 & 1.74589 &   H I 1215   &  &\\
  3347.052s&    13.555 $\pm$ 0.04 &     23.69$\pm$ 2.53 & 1.75326 &   H I 1215   & &  \\
  3348.434s&    13.665 $\pm$ 0.04 &     28.67$\pm$ 2.59 & 1.75439 &   H I 1215   & &  \\
  3350.560 &    13.569 $\pm$ 0.03 &     31.76$\pm$ 3.26 & 1.75614 &   H I 1215   & &  \\
  3351.220 &    13.760 $\pm$ 0.16 &     10.36$\pm$ 3.13 & 1.75669 &   H I 1215   & &  \\
  3351.785 &    13.830 $\pm$ 0.04 &     28.62$\pm$ 4.21 & 1.75715 &   H I 1215   & &  \\
  3352.324 &    13.230 $\pm$ 0.07 &     37.60$\pm$ 8.50 & 1.75759 &   H I 1215   &  &  \\
  3356.996s&    13.958 $\pm$ 0.02 &     28.30$\pm$ 0.95 & 1.76144 &   H I 1215   & &  \\
  3370.170 &    13.658 $\pm$ 0.10 &     17.65$\pm$ 0.01 & 1.79334 &   Si III 1206&  &\\
  3370.677 &    13.068 $\pm$ 0.15 &     15.00$\pm$ 0.03 & 1.79377 &   Si III 1206&  &\\
  3371.046 &    13.839 $\pm$ 0.40 &     13.00$\pm$ 1.94 & 1.79407 &   Si III 1206 & &\\
  3371.852 &    13.719 $\pm$ 0.09 &     29.27$\pm$ 6.24 & 1.79474 &   Si III 1206 & &\\
  3372.446 &    15.618 $\pm$ 0.17 &      4.36$\pm$ 9.08 & 1.79523 &   Si III 1206 & &\\
  3372.882 &    12.559 $\pm$ 0.13 &     15.38$\pm$ 9.26 & 1.79559 &   Si III 1206 & &\\
  3373.253 &    12.562 $\pm$ 0.10 &      8.59$\pm$ 13.18 & 1.79590 &   Si III 1206 & &\\
  3373.671 &    12.480 $\pm$ 0.09 &     13.77$\pm$ 16.61 & 1.79625 &   Si III 1206 & &\\
  3382.200b&    13.493 $\pm$ 0.06 &    109.74$\pm$20.84 & 1.78217 &   H I 1215    & &\\
  3384.814s&    13.318 $\pm$ 0.05 &     19.36$\pm$ 3.54 & 1.78432 &   H I 1215    & & \\
  3392.583b&    14.279 $\pm$ 0.05 &     43.05$\pm$ 2.95 & 1.79071 &   H I 1215  &&\\
  3393.472b&    14.241 $\pm$ 0.41 &     12.73$\pm$ 4.94 & 1.79144 &   H I 1215  && \\
  3394.502b&    13.611 $\pm$ 0.05 &     36.35$\pm$ 5.19 & 1.79229 &   H I 1215  && \\
  3395.345 &    14.655 $\pm$ 0.32 &     23.73$\pm$ 5.75 & 1.79298 &   A H I 1215 & & \\
  3396.222 &    14.975 $\pm$ 0.26 &     10.00$\pm$ 0.00 & 1.79370 &   A H I 1215 & & \\
  3396.722 &    15.446 $\pm$ 0.46 &     10.00$\pm$ 0.00 & 1.79412 &   A H I 1215 & & \\
  3397.422 &    17.386 $\pm$ 0.32 &     10.00$\pm$ 0.00 & 1.79469 &   A H I 1215 & & \\
  3397.702 &    11.938 $\pm$ 3.40 &     10.00$\pm$ 0.00 & 1.79492 &   A H I 1215 & & \\
  3398.062 &    16.394 $\pm$ 0.79 &     10.00$\pm$ 0.00 & 1.79522 &   A H I 1215 & & \\
  3398.552 &    13.447 $\pm$ 0.55 &     10.00$\pm$ 0.00 & 1.79562 &   A H I 1215 & & \\
  3399.152 &    18.364 $\pm$ 0.09 &     10.00$\pm$ 0.00 & 1.79611 &   A H I 1215 & & \\
  3400.564 &    16.583 $\pm$ 0.52 &      7.81$\pm$ 0.41 & 1.79728 &   A H I 1215 & & \\
  3401.172 &    13.447 $\pm$ 0.16 &     13.95$\pm$ 3.40 & 1.79778 &   A H I 1215 & & \\
\hline
\end{tabular}
\end{table}
\begin{table}
\begin{tabular}{lcrcrll} \hline
\multicolumn {1}{c}{$\lambda$}&\multicolumn{1}{c}{log N(X)}&
\multicolumn {1}{c}{b (km s$^{-1}$)}&\multicolumn{1}{c}{z}&
\multicolumn {1}{c}{ID, X}&\multicolumn {1}{c}{}&\multicolumn {1}{c}{Remarks}\\
\hline
  3414.694s&    13.684 $\pm$ 0.03 &     27.86$\pm$ 1.79 & 1.80890 &   H I 1215    & *& \\
  3430.562 &    14.301 $\pm$ 0.06 &     34.63$\pm$ 4.15 & 1.82195 &   H I 1215    & *& \\
  3431.239 &    13.914 $\pm$ 0.10 &     78.55$\pm$11.06 & 1.82251 &   H I 1215    & *& \\
  3432.236 &    12.969 $\pm$ 0.10 &      6.45$\pm$ 5.73 & 1.82333 &   H I 1215    &  & \\
  3432.766 &    13.418 $\pm$ 0.05 &     33.39$\pm$ 4.21 & 1.82376 &   H I 1215    & *& \\
  3443.481 &    13.161 $\pm$ 0.09 &     19.42$\pm$ 6.71 & 1.83258 &   H I 1215    & & \\
  3444.282 &    13.879 $\pm$ 0.06 &     25.89$\pm$ 3.30 & 1.83324 &   H I 1215    & *& \\
  3445.317 &    13.592 $\pm$ 0.10 &     21.24$\pm$ 8.02 & 1.22538 &   C IV 1548   & &\\
  3449.274 &    13.650 $\pm$ 0.07 &     19.17$\pm$ 3.32 & 1.83734 &   H I 1215    & *& \\
  3449.827 &    13.158 $\pm$ 0.10 &     38.70$\pm$ 3.51 & 1.83780 &   H I 1215    &  & \\
  3451.048 &    13.592 $\pm$ 0.04 &     21.24$\pm$ 3.51 & 1.22538 &   C IV 1550   & &\\
  3454.151s&    13.217 $\pm$ 0.05 &     18.16$\pm$ 2.82 & 1.84136 &   H I 1215    & *& \\
  3454.964s&    13.364 $\pm$ 0.04 &     17.08$\pm$ 2.01 & 1.84202 &   H I 1215    & *& \\
  3458.571s&    13.344 $\pm$ 0.03 &     31.05$\pm$ 3.03 & 1.84499 &   H I 1215    & *& \\
  3462.256 &    13.785 $\pm$ 0.05 &     16.12$\pm$ 2.68 & 1.79480 &   N V 1238    & &\\
  3462.875d&    13.549 $\pm$ 0.04 &     33.16$\pm$ 3.84 & 1.84853 &   H I 1215    & *& \\
  3465.643s&    13.013 $\pm$ 0.05 &     16.73$\pm$ 3.62 & 1.85081 &   H I 1215    &  & \\
  3473.377 &    13.785 $\pm$ 0.05 &     16.12$\pm$ 2.68 & 1.79480 &   N V 1238    & &\\
  3478.466s&    13.498 $\pm$ 0.02 &     21.98$\pm$ 0.98 & 1.86136 &   H I 1215    &  *&\\
  3499.087s&    13.111 $\pm$ 0.03 &     19.84$\pm$ 1.83 & 1.87832 &   H I 1215  &    & \\
  3503.030 &    13.121 $\pm$ 0.06 &     41.76$\pm$ 7.37 & 1.88156 &   H I 1215 &   & Po B-C II $\lambda$1334\\
  3503.752 &    12.924 $\pm$ 0.06 &     14.67$\pm$ 3.41 & 1.88216 &   H I 1215 &  & Po B-C II $\lambda$1334\\
  3507.435 &    17.389 $\pm$ 0.17 &      7.77$\pm$ 0.18 & 1.88519 &   C H I 1215 & & \\
  3508.325 &    14.310 $\pm$ 0.20 &     60.82$\pm$36.56 & 1.88592 &   C H I 1215  & &\\
  3508.903 &    13.977 $\pm$ 1.36 &      6.57$\pm$36.44 & 1.88639 &   C H I 1215  & &\\
  3509.399 &    14.000 $\pm$ 0.18 &     32.30$\pm$ 5.91 & 1.88680 &   C H I 1215  & &\\
  3511.933b&    13.986 $\pm$ 0.02 &    105.74$\pm$ 7.01 & 1.88889 &   C H I 1215  &£&\\
  3518.631s&    13.728 $\pm$ 0.01 &     22.30$\pm$ 0.58 & 1.89440 &   H I 1215    & *& \\
  3519.643 &    13.947 $\pm$ 0.03 &     14.51$\pm$ 0.64 & 1.89523 &   D H I 1215  & &\\
  3520.064 &    13.857 $\pm$ 0.08 &     14.47$\pm$ 3.57 & 1.89558 &   D H I 1215  & &\\
  3520.643 &    14.515 $\pm$ 0.10 &     21.32$\pm$ 2.28 & 1.89605 &   D H I 1215  & &\\
  3521.572 &    13.776 $\pm$ 0.05 &     31.92$\pm$ 5.36 & 1.79372 &   Si II 1260  & &\\
  3522.078 &    13.228 $\pm$ 0.10 &     11.33$\pm$ 2.02 & 1.79412 &   Si II 1260  & &\\
  3522.860 &    13.929 $\pm$ 0.04 &     13.74$\pm$ 0.48 & 1.79474 &   Si II 1260  & &\\
  3523.125 &    13.262 $\pm$ 0.07 &     10.71$\pm$ 1.43 & 1.79495 &   Si II 1260  & &\\
  3523.481 &    13.068 $\pm$ 0.05 &      8.93$\pm$ 1.23 & 1.79523 &   Si II 1260  & &\\
  3523.936 &    12.964 $\pm$ 0.03 &     21.89$\pm$ 1.70 & 1.79559 &   Si II 1260  & &\\
  3524.586 &    12.442 $\pm$ 0.05 &     15.56$\pm$ 2.83 & 1.79611 &   Si II 1260  & &\\
  3532.849 &    14.061 $\pm$ 0.03 &     38.37$\pm$ 2.38 & 1.90609 &   H I 1215    & *& \\
  3533.448 &    13.407 $\pm$ 0.12 &      8.43$\pm$ 5.72 & 1.90658 &   H I 1215    & *& \\
  3533.925 &    13.942 $\pm$ 0.04 &     25.77$\pm$ 2.23 & 1.90698 &   H I 1215    & *& \\
  3543.849s&    13.422 $\pm$ 0.03 &     44.72$\pm$ 4.68 & 1.91514 &   H I 1215 & *& PC H-Fe II$\lambda$2600\\
  3548.555s&    13.104 $\pm$ 0.02 &     13.96$\pm$ 1.34 & 1.91901 &   H I 1215    &  & \\
  3553.827s&    13.188 $\pm$ 0.07 &     27.78$\pm$ 6.74 & 1.92335 &   H I 1215    &  & \\
  3570.644s&    13.626 $\pm$ 0.02 &     27.33$\pm$ 1.41 & 1.93718 &   H I 1215    & *& \\
  3573.287s&    13.588 $\pm$ 0.01 &     23.44$\pm$ 0.76 & 1.93936 &   H I 1215    & *& \\
  3574.467s&    12.889 $\pm$ 0.02 &     16.15$\pm$ 1.28 & 1.94033 &   H I 1215    & & \\
  3577.160b&    13.009 $\pm$ 0.06 &     27.06$\pm$ 6.32 & 1.94254 &   H I 1215 & & \\
  3596.782 &    13.093 $\pm$ 0.04 &     22.00$\pm$ 3.41 & 1.95868 &   H I 1215 &   &  \\
  3597.538 &    13.258 $\pm$ 0.03 &     19.89$\pm$ 2.30 & 1.95930 &   H I 1215 &  * &  \\
  3604.661 &    14.581 $\pm$ 0.06 &     34.42$\pm$ 1.35 & 1.96516 &   H I 1215 & *  &  \\
  3605.735 &    13.137 $\pm$ 0.04 &      7.90$\pm$ 1.54 & 1.96605 &   H I 1215 &    &  \\
  3607.005b&    13.199 $\pm$ 0.08 &     44.66$\pm$13.07 & 1.96709 &   H I 1215 &    &\\
  3611.744s&    12.884 $\pm$ 0.06 &     15.25$\pm$ 3.65 & 1.97099 &   H I 1215 &    &  \\
  3617.934s&    13.927 $\pm$ 0.01 &     35.06$\pm$ 0.95 & 1.97608 &   H I 1215 & *  & \\
  3621.071 &    13.161 $\pm$ 0.03 &     12.42$\pm$ 1.70 & 1.97866 &   H I 1215 &    &  \\
  3621.701 &    13.913 $\pm$ 0.02 &     25.50$\pm$ 1.47 & 1.97918 &   H I 1215 & *  &  \\
  3622.341 &    13.688 $\pm$ 0.03 &     18.18$\pm$ 1.82 & 1.97971 &   H I 1215 & *  &  \\
  3622.896 &    13.720 $\pm$ 0.03 &     22.11$\pm$ 1.71 & 1.98016 &   H I 1215 & *  &  \\
  3624.112b&    13.885 $\pm$ 0.03 &     72.80$\pm$ 5.84 & 1.98116 &   H I 1215   & ? & \\
  3625.667 &    14.250 $\pm$ 0.02 &     46.14$\pm$ 1.17 & 1.98244 &   H I 1215    & *& \\
  3633.688s&    13.378 $\pm$ 0.03 &     18.74$\pm$ 2.00 & 1.98904 &   H I 1215     & *& \\
\hline
\end{tabular}
\end{table}
\begin{table}
\begin{tabular}{lcrcrll} \hline
\multicolumn {1}{c}{$\lambda$}&\multicolumn{1}{c}{log N(X)}&
\multicolumn {1}{c}{b (km s$^{-1}$)}&\multicolumn{1}{c}{z}&
\multicolumn {1}{c}{ID, X}&\multicolumn {1}{c}{}&\multicolumn {1}{c}{Remarks}\\
\hline
  3637.900 &    14.396 $\pm$ 0.05 &     23.50$\pm$ 2.16 & 1.79373 &   O I 1302     & &\\
  3638.136d&    13.584 $\pm$ 0.07 &     14.73$\pm$ 4.82 & 1.99270 &   H I 1215     &  *&\\
  3638.423 &    13.766 $\pm$ 0.12 &      9.73$\pm$ 3.47 & 1.79413 &   O I 1302     & &\\
  3641.228 &    14.288 $\pm$ 0.05 &     27.46$\pm$ 1.82 & 1.99524 &   H I 1215     & *& \\
  3641.864 &    13.326 $\pm$ 0.14 &     22.41$\pm$ 5.21 & 1.99577 &   H I 1215     & *& \\
  3644.041 &    13.776 $\pm$ 0.05 &     31.92$\pm$ 4.53 & 1.79372 &   Si II 1304   & &\\
  3644.380d&    13.544 $\pm$ 0.05 &     24.18$\pm$ 1.92 & 1.99784 &   H I 1215     & *& \\
  3644.563 &    13.228 $\pm$ 0.18 &     11.33$\pm$ 9.83 & 1.79412 &   Si II 1304   & &\\
  3645.373 &    13.929 $\pm$ 0.09 &     13.74$\pm$ 2.62 & 1.79474 &   Si II 1304   & &\\
  3645.460d&    13.428 $\pm$ 0.06 &     37.60$\pm$ 4.35 & 1.99872 &   H I 1215     & *& \\
  3645.648 &    13.262 $\pm$ 0.22 &     10.71$\pm$ 8.89 & 1.79495 &   Si II 1304   & &\\
  3646.016 &    13.068 $\pm$ 0.27 &      8.93$\pm$ 7.77 & 1.79523 &   Si II 1304   & &\\
  3646.486 &    12.964 $\pm$ 0.71 &     21.89$\pm$ 13.78 & 1.79559 &   Si II 1304   & &\\
  3647.159 &    12.442 $\pm$ 0.12 &     15.56$\pm$ 27.95 & 1.79611 &   Si II 1304   & &\\
  3651.995s&    13.975 $\pm$ 0.03 &     28.87$\pm$ 1.05 & 2.00410 &   H I 1215     & *& \\
  3657.067s&    13.025 $\pm$ 0.02 &     20.20$\pm$ 1.59 & 2.00827 &   H I 1215     &  & \\
  3659.730b&    13.100 $\pm$ 0.05 &     44.37$\pm$ 7.69 & 2.01046 &   H I 1215     &  &\\
  3662.339 &    13.561 $\pm$ 0.11 &     23.72$\pm$ 3.50 & 2.01261 &   H I 1215     & *& \\
  3662.890 &    13.750 $\pm$ 0.08 &     33.38$\pm$ 4.49 & 2.01306 &   H I 1215     & *& \\
  3670.743 &    13.037 $\pm$ 0.23 &     33.11$\pm$16.06 & 2.01952 &   H I 1215     &  & \\
  3671.432 &    13.999 $\pm$ 0.04 &     26.99$\pm$ 1.93 & 2.02009 &   H I 1215    & * & \\
  3674.240s&    13.185 $\pm$ 0.02 &     20.09$\pm$ 1.23 & 2.02240 &   H I 1215     &  & \\
  3676.134s&    13.739 $\pm$ 0.02 &     30.65$\pm$ 1.73 & 2.02396 &   H I 1215     & *&  \\
  3680.083 &    13.217 $\pm$ 0.03 &     38.02$\pm$ 3.53 & 2.02721 &   H I 1215     & *& \\
  3681.644 &    14.328 $\pm$ 0.03 &     41.88$\pm$ 1.13 & 2.02849 &   H I 1215     & *& \\
  3688.010s&    12.988 $\pm$ 0.03 &     27.99$\pm$ 2.79 & 2.03373 &   H I 1215     &  & \\
  3697.779s&    14.009 $\pm$ 0.01 &     36.28$\pm$ 0.67 & 2.04176 &   H I 1215     & *& \\
  3701.355 &    13.146 $\pm$ 0.09 &     19.05$\pm$ 3.71 & 2.04470 &   H I 1215     &  & \\
  3701.994 &    14.041 $\pm$ 0.02 &     26.51$\pm$ 1.37 & 2.04523 &   H I 1215     & *& \\
  3703.485b&    12.973 $\pm$ 0.05 &     41.65$\pm$ 7.27 & 2.04646 &   H I 1215     & &\\
  3709.109 &    13.648 $\pm$ 0.08 &     35.40$\pm$ 4.84 & 2.05108 &   H I 1215     & *& \\
  3710.044 &    13.494 $\pm$ 0.14 &     42.77$\pm$17.45 & 2.05185 &   H I 1215     & *& \\
  3710.669 &    12.808 $\pm$ 0.19 &      9.84$\pm$ 7.22 & 2.05237 &   H I 1215     & & \\
  3712.725b&    13.143 $\pm$ 0.07 &     77.65$\pm$17.94 & 2.05406 &   H I 1215     & &\\
  3723.698 &    12.948 $\pm$ 0.04 &     15.72$\pm$ 2.23 & 2.06308 &   H I 1215     &  & \\
  3725.157b&    13.255 $\pm$ 0.03 &     66.35$\pm$ 7.13 & 2.06428 &   H I 1215     &?&\\
  3727.675 &    14.124 $\pm$ 0.02 &     22.89$\pm$ 1.07 & 1.79324 &   C II 1334    & &\\
  3727.989 &    14.270 $\pm$ 0.04 &     21.38$\pm$ 2.15 & 1.79348 &   C II 1334    & &\\
  3728.671 &    14.754 $\pm$ 0.03 &     24.36$\pm$ 0.83 & 1.79399 &   C II 1334    & &\\
  3729.403 &    14.210 $\pm$ 0.04 &      9.36$\pm$ 0.81 & 1.79454 &   C II 1334    & &\\
  3729.899 &    16.193 $\pm$ 0.09 &      8.95$\pm$ 0.31 & 1.79491 &   C II 1334    & &\\
  3730.348 &    14.182 $\pm$ 0.03 &     16.92$\pm$ 1.31 & 1.79525 &   C II 1334    & &\\
  3730.836 &    13.957 $\pm$ 0.02 &     16.23$\pm$ 1.08 & 1.79561 &   C II 1334    & &\\
  3731.344 &    13.477 $\pm$ 0.03 &      9.53$\pm$ 1.65 & 1.79599 &   C II 1334    & &\\
  3739.857s&    12.920 $\pm$ 0.05 &     22.11$\pm$ 4.07 & 2.07638 &   H I 1215     &  & \\
  3742.297s&    13.396 $\pm$ 0.03 &     22.72$\pm$ 2.08 & 2.07838 &   H I 1215     & *& \\
  3743.280s&    13.632 $\pm$ 0.01 &     24.23$\pm$ 0.61 & 2.07919 &   H I 1215     & *& \\
  3745.524b&    14.190 $\pm$ 0.03 &     32.30$\pm$ 1.79 & 2.08104 &   H I 1215     &?& \\
  3746.846b&    14.068 $\pm$ 0.02 &     39.09$\pm$ 1.97 & 2.08212 &   H I 1215     &?& \\
  3751.727b&    13.364 $\pm$ 0.03 &     46.03$\pm$ 4.59 & 2.08614 &   H I 1215     &?&\\
  3756.656b&    13.548 $\pm$ 0.03 &     53.98$\pm$ 4.78 & 2.09019 &   H I 1215  &?& PC C-O I $\lambda$1302\\
  3758.338s&    13.207 $\pm$ 0.02 &     18.35$\pm$ 1.24 & 2.09158 &   H I 1215  & & Po C-O I $\lambda$1302\\
  3759.390 &    13.615 $\pm$ 0.03 &     18.07$\pm$ 1.94 & 1.42824 &   C IV 1548  & & \\
  3760.124 &    13.790 $\pm$ 0.03 &     15.50$\pm$ 1.59 & 1.42871 &   C IV 1548   &&  \\
  3760.585 &    13.204 $\pm$ 0.07 &     16.61$\pm$ 4.35 & 1.42901 &   C IV 1548    && \\
  3763.183s&    13.386 $\pm$ 0.04 &     15.24$\pm$ 1.88 & 2.09556 &   H I 1215     & *& \\
  3765.643 &    13.615 $\pm$ 0.03 &     18.07$\pm$ 1.94 & 1.42824 &   C IV 1550    & &\\
  3766.378 &    13.790 $\pm$ 0.03 &     15.50$\pm$ 1.59 & 1.42871 &   C IV 1550    & &\\
  3766.840 &    13.204 $\pm$ 0.07 &     16.61$\pm$ 4.35 & 1.42901 &   C IV 1550  &  &\\
  3771.969s&    13.701 $\pm$ 0.02 &     33.49$\pm$ 1.87 & 2.10279 &   H I 1215  & *& \\
  3776.511s&    12.912 $\pm$ 0.03 &      5.35$\pm$ 0.82 & 2.10653 &   H I 1215   & &   \\
  3779.612s&    13.878 $\pm$ 0.01 &     41.34$\pm$ 0.99 & 2.10908 &   H I 1215    & *&  \\
\hline
\end{tabular}
\end{table}
\begin{table}
\begin{tabular}{lcrcrll} \hline
\multicolumn {1}{c}{$\lambda$}&\multicolumn{1}{c}{log N(X)}&
\multicolumn {1}{c}{b (km s$^{-1}$)}&\multicolumn{1}{c}{z}&
\multicolumn {1}{c}{ID, X}&\multicolumn {1}{c}{}&\multicolumn {1}{c}{Remarks}\\
\hline
  3790.585 &    13.571 $\pm$ 0.11 &     24.49$\pm$ 6.56 & 2.11810 &   G H I 1215   & &\\
  3791.145 &    14.498 $\pm$ 0.03 &     57.69$\pm$ 2.62 & 2.11856 &   G H I 1215   & &\\
  3792.781 &    14.642 $\pm$ 0.10 &     33.69$\pm$ 2.01 & 2.11991 &   G H I 1215   & &\\
  3799.107 &    13.591 $\pm$ 0.01 &     31.75$\pm$ 0.98 & 2.12511 &   H I 1215     & *& \\
  3799.942 &    12.960 $\pm$ 0.03 &     16.92$\pm$ 1.66 & 2.12580 &   H I 1215     &  & \\
  3802.694s&    13.676 $\pm$ 0.02 &     37.39$\pm$ 1.60 & 2.12806 &   H I 1215     & *& \\
  3805.392b&    13.068 $\pm$ 0.06 &     37.69$\pm$ 6.85 & 2.13028 &   H I 1215     & &\\
  3808.647s&    12.897 $\pm$ 0.03 &     14.69$\pm$ 2.02 & 2.13296 &   H I 1215     &  & \\
  3811.431d&    15.072 $\pm$ 0.24 &     31.64$\pm$ 3.83 & 2.13525 &   H I 1215     & *& \\
  3811.525 &    13.481 $\pm$ 0.03 &     52.37$\pm$ 2.80 & 0.36303 &   Mg II 2796   & &\\
  3812.014 &    13.759 $\pm$ 0.04 &     26.60$\pm$ 1.24 & 0.36321 &   Mg II 2796   & &\\
  3817.186 &    13.100 $\pm$ 0.04 &     31.83$\pm$ 4.03 & 2.13998 &   H I 1215     &  & \\
  3821.311 &    13.481 $\pm$ 0.03 &     52.37$\pm$ 2.80 & 0.36303 &   Mg II 2803   & &\\
  3821.800 &    13.759 $\pm$ 0.04 &     26.60$\pm$ 1.24 & 0.36321 &   Mg II 2803   & &\\
  3822.672d&    13.900 $\pm$ 0.02 &     30.41$\pm$ 1.47 & 2.14450 &   H I 1215     & *& \\
  3826.475s&    12.733 $\pm$ 0.03 &     18.45$\pm$ 2.41 & 2.14763 &   H I 1215     &  & \\
  3834.612s&    12.934 $\pm$ 0.05 &     16.86$\pm$ 3.07 & 2.15432 &   H I 1215     &  & \\
  3838.264b&    12.971 $\pm$ 0.07 &     40.36$\pm$ 8.93 & 2.15732 &   H I 1215     & &\\
  3839.764b&    13.057 $\pm$ 0.10 &     49.57$\pm$15.76 & 2.15856 &   H I 1215     & &\\
  3844.764b&    13.305 $\pm$ 0.04 &     55.13$\pm$ 5.47 & 2.16267 &   H I 1215     &?&\\
  3850.580s&    12.843 $\pm$ 0.06 &     20.94$\pm$ 4.55 & 2.16745 &   H I 1215  &   &PC C-C II $\lambda$1334\\
  3852.793b&    13.220 $\pm$ 0.04 &     51.98$\pm$ 6.72 & 2.16928 &   H I 1215  &?&PC C-C II $\lambda$1334\\
  3857.335 &    13.013 $\pm$ 0.03 &      9.84$\pm$ 1.88 & 2.17301 &   H I 1215   & &   \\
  3858.039 &    13.049 $\pm$ 0.04 &     24.85$\pm$ 3.03 & 2.17359 &   H I 1215    &  & \\
  3863.990b&    13.204 $\pm$ 0.16 &     94.36$\pm$47.07 & 2.17849 &   H I 1215&?&PC D-C II $\lambda$1334 \&\\
           &                      &                     &         &           & &  G-N V $\lambda$1238\\
  3865.848b&    13.068 $\pm$ 0.06 &     50.91$\pm$ 9.74 & 2.18001 &   H I 1215  & & PC D-C II$\lambda$1334 \&\\
           &                      &                     &         &             & & G-N V $\lambda$1238\\
  3890.501s&    13.009 $\pm$ 0.03 &      8.85$\pm$ 1.72 & 2.20029 &   H I 1215     & & \\
  3893.178 &    13.158 $\pm$ 0.08 &     12.40$\pm$ 3.44 & 1.79330 &   Si IV 1393   & &\\
  3893.888 &    13.885 $\pm$ 0.14 &     24.81$\pm$ 14.81 & 1.79381 &   Si IV 1393   & &\\
  3894.166d&    14.522 $\pm$ 0.13 &     35.61$\pm$ 10.73 & 2.20331 &   H I 1215     & & \\
  3894.351 &    13.274 $\pm$ 2.11 &      6.54$\pm$ 38.13 & 1.79414 &   Si IV 1393   & &\\
  3895.073 &    13.769 $\pm$ 0.15 &     10.66$\pm$ 7.01 & 1.79466 &   Si IV 1393   & &\\
  3895.488 &    14.076 $\pm$ 0.20 &     17.40$\pm$ 4.80 & 1.79496 &   Si IV 1393   & &\\
  3895.970d&    13.725 $\pm$ 0.16 &      7.80$\pm$ 2.89 & 2.20479 &   H I 1215     & & \\
  3896.307 &    13.342 $\pm$ 0.08 &     20.45$\pm$ 3.54 & 1.79555 &   Si IV 1393   & &\\
  3897.053 &    13.004 $\pm$ 0.08 &     23.73$\pm$ 3.27 & 1.79608 &   Si IV 1393   & &\\
  3900.517s&    13.064 $\pm$ 0.04 &     23.20$\pm$ 3.47 & 2.20853 &   H I 1215     & & \\
  3905.918b&    13.086 $\pm$ 0.06 &     64.02$\pm$12.06 & 2.21298 &   H I 1215     & &\\
  3907.887b&    13.053 $\pm$ 0.06 &     36.65$\pm$ 6.68 & 2.21460 &   H I 1215     & &\\
  3916.103b&    12.827 $\pm$ 0.06 &     44.20$\pm$ 9.30 & 2.22135 &   H I 1215     & &\\
  3918.360 &    13.158 $\pm$ 0.08 &     12.40$\pm$ 3.44 & 1.79330 &   Si IV 1402   & &\\
  3919.074 &    13.885 $\pm$ 0.14 &     24.81$\pm$ 14.81 & 1.79381 &   Si IV 1402   & &\\
  3919.541 &    13.274 $\pm$ 2.11 &      6.54$\pm$ 38.13 & 1.79414 &   Si IV 1402   & &\\
  3920.267 &    13.769 $\pm$ 0.15 &     10.66$\pm$ 7.01 & 1.79466 &   Si IV 1402   & &\\
  3920.685 &    14.076 $\pm$ 0.20 &     17.40$\pm$ 4.80 & 1.79496 &   Si IV 1402   & &\\
  3921.509 &    13.342 $\pm$ 0.08 &     20.45$\pm$ 3.54 & 1.79555 &   Si IV 1402   & &\\
  3922.259 &    13.004 $\pm$ 0.08 &     23.73$\pm$ 3.27 & 1.79608 &   Si IV 1402   & &\\
  3938.544s&    13.158 $\pm$ 0.04 &     30.21$\pm$ 4.01 & 2.23981 &   H I 1215     & & \\
\hline
\end{tabular}

 b Line shape is not proper, probably because of blends or bad pixel\\
 s Lines fitted with single components\\
 d Lyman alpha forest lines which are deblended from metal lines\\
 $*$ Lines included in the Lyman alpha line sample \\
 ? Lines that can be included in the Lyman alpha line sample but for their 'unusual' shape\\
Po Possibly\\
Co Contaminated by\\ 
PC Possibly contaminated by\\
\end{table}
\vfil\eject
\begin{table}
\caption {Column density distribution }
\bigskip
\begin{tabular}{cccccc} 
\hline
\multicolumn{1}{c}{QSO}&
\multicolumn {1}{c}{log N$_{\rm H\; I}^{min}$}&\multicolumn{1}{c}{No.of Lines}&
\multicolumn{1}{c}{$\beta$}&\multicolumn{1}{c}{$<$ z $>$}&\multicolumn {1}{c}{P$_{\rm ks}$}\\
\hline
B2 1225+317 & 13.2 & 52 & 1.828 $\pm 0.115$ & 1.9768 & 0.088 \\
 & 13.4 & 42 & 2.025 $\pm 0.158$ & 1.9768 & 0.337 \\
 & 13.6 & 30 & 2.241 $\pm 0.226$ & 1.9851 & 0.526 \\
 & 13.8 & 18 & 2.334 $\pm 0.314$ & 1.9843 & 0.387 \\
 & 14.0 &  9 & 2.333 $\pm 0.444$ & 1.9912 & 0.854 \\
 & & & & & \\
B2 1225+317 & 13.2 & 190 & 1.742 $\pm 0.054$ & 2.0195 & 0.039 \\
Q1331+170& 13.4 & 147 & 1.829 $\pm 0.068$ & 2.0226 & 0.113 \\
 Q1101-26& 13.6 & 110 & 1.938 $\pm 0.089$ & 2.0304 & 0.174 \\
 \&  & 13.8 & 80 & 2.080 $\pm 0.121$ & 2.0444 & 0.575 \\
Q2206-199 & 14.0 & 45 & 2.014 $\pm 0.151$ &2.0470 & 0.685 \\
\hline
\end{tabular}
\end{table}
\end{document}